# Atomic Distributions in Topological Insulator $Bi_2Se_{3-x}Te_x$

Kaixin Huang[1,*], Haotian Zheng[1,*], and Weidong Luo[1,2,3,†]

[1]Department of Physics and Astronomy, Shanghai Jiao Tong University, Shanghai 200240, China

[2] Institute of Natural Sciences, Shanghai Jiao Tong University, Shanghai 200240, China

[3] Collaborative Innovation Center of Advanced Microstructures, Nanjing 210046, China

$Bi_2Se_3$ is a topological insulator and it is often doped with Te to compensate the *n*-type carriers due to Se vacancies. Different doping patterns of Te would influence the transport characteristics of the surface states. We study the Te atom distribution in $Bi_2Se_3$ with different Te concentrations, using first-principles density-functional-theory calculations. We show that Te prefer the outer layer, until the composition becomes $Bi_2SeTe_2$ and the outer layer is full of Te. And for a fixed ratio of Se and Te atoms within a single layer, we find that the structures with less number of adjacent Te-Se pairs tend to have lower energies. This might result in the formation of local Te clusters under low annealing temperatures.

**Introduction**

Topological insulator is a new state of matter with fascinating properties [14][15], and it is one of the research hotspots of physics in recent years. It differs from traditional metal and insulator, as it is insulating inside the material, while the surface or interface allows charge transport. But in reality the common topological insulators such as $Bi_2Se_3$ often has internal defects, such as Se vacancies, and the material is *n*-type doped. In this case the surface states contribution of transport property is smaller than the bulk electronic state, by 1-2 orders of magnitude. As a result, the material does not show the expected transport characteristics of a topological insulator. The incorporation of Te into $Bi_2Se_3$ introduces *p*-type carriers, mainly anti-site defects such as Bi on the Te sites, which compensates the original *n*-type carriers of $Bi_2Se_3$, thereby moving the Fermi level to the bulk band gap and reducing the contribution of bulk electron on transport property[1][2][5][6]. Depending on the Te concentrations, we often encounter ordered or disordered Se and Te structures. Powder XRD experiments revealed that Te atoms in $Bi_2Te_2Se$ fill in the outer side and Se atoms fill in the inner side [17]. It has been shown that doped Te prefers the outer side (site (I)) of $Bi_2Se_3$, as shown in fig.1.

In the present study, we use the statistical random sampling method combined with first-principles density-functional-theory calculations to study the Se and Te distributions. We obtain similar trend for the positions of the Te atoms in $Bi_2Se_{3-x}Te_x$, and related materials. We develop a model based on the number of adjacent Se-Te pairs, which can explain the doping trend as well as the total energy dependence on Te doping.

**Method**

This work is based on first-principles density-functional-theory (DFT) calculations as implemented in the Vienna ab initio simulation package (VASP)[Error! Reference source not found.][9][10]. We use atomic potentials from the projector augmented-wave (PAW)[11][12] method, and exchange-correlation functional in the generalized-gradient approximation parameterized by Perdew, Burke, and Ernzerhof (GGA-PBE)[7][8]. And after some tests we choose the k-point samples with $2 \times 2 \times 1$ for both monolayers and bulk geometries in their $4 \times 4 \times 1$ supercell structure. All geometries, except the lattice constant on the *z*-axis in monolayer calculations, are relaxed until the residual forces are no larger than $10^{-3}$ eV/Å. The calculations incorporate the van der Waals interaction with the optB86b-vdW approximation [13]. In the calculations of monolayer we use a vacuum spacing of 30 Å along the *z*-direction to avoid spurious interactions.

**Results**

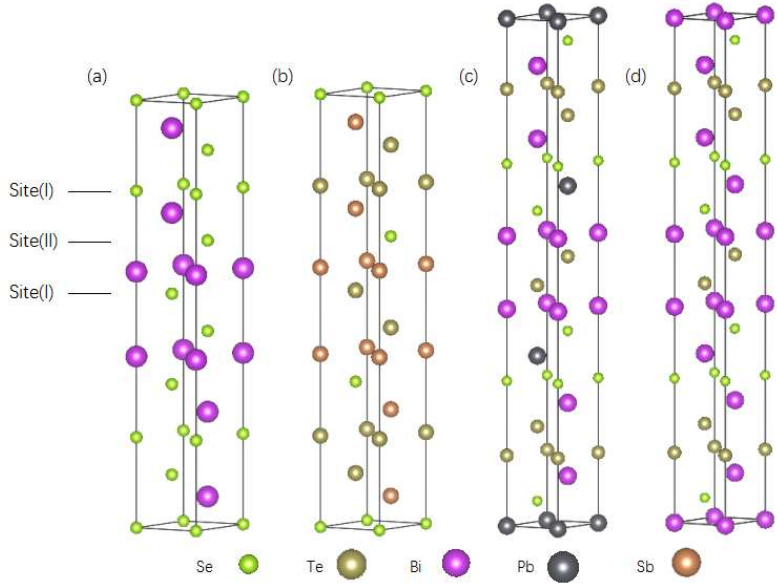

FIG. 1. (a) $Bi_2Se_3$ crystallographic cell with hexagonal symmetry, (b) $Bi_2PbSe_2Te_2$ crystallographic cell, (c) $Bi_3Se_4$ crystallographic cell, (d) $Sb_2Te_2Se$ crystallographic cell.

$Bi_2Se_3$ is an isostructural five-layer tetradymite-type compounds, in which the layers stack in the sequence Se-Bi-Se-Bi-Se. One crystallographic cell includes three five-layer tetradymite-type units stacked with hexagonal symmetry. Each five-layer unit (quintuple layer, QL) interacts with other units by van der Waals interaction between the outer layer Se atoms of the QL. We define the outer site Se as Se(I), and the inner site Se as Se(II), as shown in fig. 1 (a). The Se atoms in site (I) and site (II) bond to three and six nearest-neighbors Bi atoms, respectively. Our calculations are based on the experimental crystal structure of $Bi_2Se_3$. We use first-principles DFT calculations to obtain the total energy of $Bi_2Se_3$ and related compounds such as $Bi_2Te_3$ and $Bi_2Se_2Te$.

In order to investigate the structural configurations of Te, in the Te-doped $Bi_2Se_3$, we use a $4 \times 4 \times 1$ supercell of the original hexagonal unit cell and we set up Te atoms substituting Se atoms at random positions. For $Bi_2Se_2Te$, we constrain Te atoms in the outer two layers (site I) and construct 10 random structures. After structural relaxation, the average energy is set to be zero as the reference, and the calculated energy standard deviation is 0.023 eV. In comparison, we also construct 10 random structures in which Te atoms are distributed in both the inner layer and the outer layers., The average energy of the supercell becomes 3.11 eV, with a standard deviation of 0.21 eV. For the ordered $Bi_2Se_2Te$ structure where all Te atoms are in the inner layer (site II), the energy is 7.53 eV. In conclusion, the average energy of the $Bi_2Se_2Te$ with Te atoms in site(I) is lower than the energy of the structure with random Te positions Te atoms doped into $Bi_2Se_3$ would first fill in the outer layer (site I).

We also consider other compounds with similar structure as $Bi_2Se_3$, and their energies are shown in table 1. Similar property is observed: larger chalcogen atoms have priority to fill in the outer layer site[16].

| Position of the larger chalcogen | $Sb_2Te_2Se$ | $Bi_2PbSe_2Te_2$ | $Bi_3Te_2Se_2$ | $Bi_3Te_2S_2$ | $Bi_3S_2Se_2$ |
|---|---|---|---|---|---|
| Outer site (eV) | -0.48 | -0.80 | -1.03 | -0.44 | -0.13 |
| Inner site (eV) | 0 | 0 | 0 | 0 | 0 |

Table 1. The energy of some related compounds, with the positions of the larger chalcogen (Te or Se) constrained in the outer site or the inner site, with the energy of the inner site structure set as the reference.,.

In order to better understand this behavior, we perform calculations in the following procedure. First, we start with the ordered structure of $Bi_2SeTe_2$, in which the Te atoms are in the outer site (site I) and the Se atoms are in the inner site (site II), and calculate the total energy of the $3 \times 3 \times 1$ hexagonal supercell. Then we randomly exchange one Te atom in site (I) with a random Se atom in site (II), and calculate the energy of the new structure. This process is repeated for further transfer of Te atoms. In the unrelaxed structure, the atoms are located in the positions of the original ordered $Bi_2SeTe_2$ structure. In the relaxed structure, the positions of all atoms are optimized. The total energies of both the unrelaxed structures and the relaxed structures are shown in fig. 2. The energies of the unrelaxed structures increase linearly with the number of Te atoms transferred from the outer layer to the inner layer. Upon structural relaxation, the energy increase is reduced and it also deviate from linear dependence.

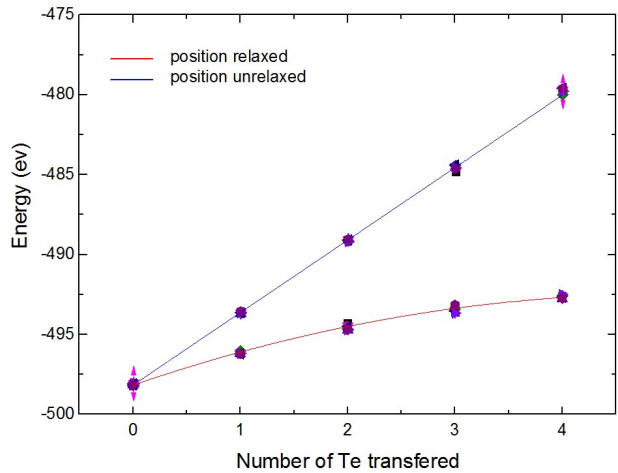

FIG. 2. Total energy of $Bi_2SeTe_2$ with Te atoms transferred from the outer layer to the inner layer. The blue line shows the energies of the unrelaxed structures in which the atoms are located in the positions of the original ordered structure, whereas the energies of the relaxed structure are shown with the red line.

We also consider the situation that Te atoms are added into $Bi_2Se_3$ to replace the Se atoms. Total energy calculations are performed for a single quintuple layer of $Bi_2Se_3$, with a $3 \times 3$ in-plane supercell. To investigate the distribution of Te in the inner layer (site II), we replace one of the Se in site(II) with Te, then perform structural relaxation and obtain the total energy. This process is repeated for additional substitution of Se with Te, and the total energy as a function of doped Te atoms is shown in fig. 3(a). If there are more than one configuration of Te and Se atoms for any particular doping of $x$ Te atoms, we construct up to 10 random configurations and calculate their total energy. For substitution of Se atoms in the outer layers (site I), similar procedure is used to obtain the total energy, also shown in fig. 3(a).

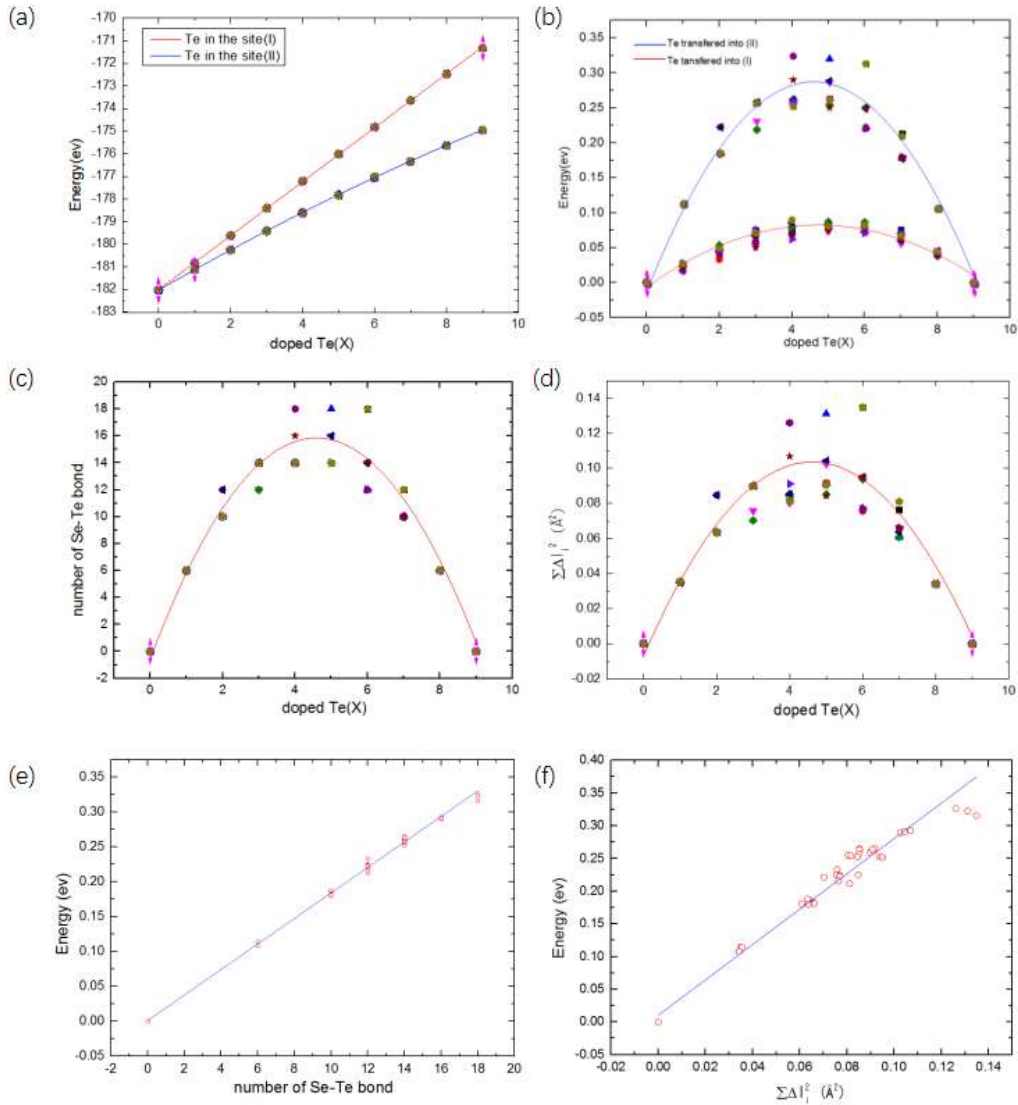

FIG. 3. (a) The total energy of $Bi_2(Se_{1-x/9}Te_{x/9})_2^I Se^{II}$ is shown with the red line where the Te atoms are doped into the outer layers (site I); the blue line represents the total energy of $Bi_2Se_2^I (Se_{1-x/9}Te_{x/9})^{II}$, where the Te atoms are doped in to the inner layer (site II); (b) the total energy of $Bi_2Se_3$ doped with Te atoms, after subtracting a linear term from the data in fig. 3(a); (c) the number of adjacent Te-Se pairs in $Bi_2Se_2^I (Se_{1-x/9}Te_{x/9})^{II}$; (d) sum of the square of the displacement of the Bi-Se and the Bi-Te bonds, calculated according to formula (4). (e) total additional energy versus the number of adjacent Te-Se pairs. (f) total additional energy versus sum of the square of the displacement of the Bi-Se bonds.

**Analysis**

A classical way to explain these doping dependence is using the electronegativity. According to the Sanderson model[3], the bonding energy of two atoms is divided into two part, the nonpolar covalent form $E_c$ and ionic form $E_i$:

$$E_c = (E_{AA}E_{BB})^{1/2} \frac{R_c}{R_0} \quad (1)$$

$$E_i = \frac{33200}{R_0} \quad (2)$$

$$E_{total} = t_c E_c + t_i E_i \quad (3)$$

Here $R_o$ is the actual bond length and $R_c$ the sum of the nonpolar covalent radii. $t_c$ and $t_i$ are covalent and ionic blending coefficients and are not known. But as the energy is linear, we expect the total energy of the $Bi_xSe_yTe_z$ system to be a combination of five parts: $E_{single\ atom}$, $E_{Bi-Se\_in}$ (which represents the bonding energy between Bi and Te in the inner layer), $E_{Bi-Se\_out}$, $E_{Bi-Te\_in}$ and $E_{Bi-Te\_out}$. The single atom energy term is usually obtained by performing calculations of a single atom placed in a large supercell. However, in this layered system, the energies of single atoms are small while the adjacent energies (between atoms in the same layer) are not negligible. Thus we actually perform calculations on a single layer of Bi, Se, or Te atoms, and obtain the sum of single atom energy and adjacent energy terms. One must caution that this is not very accurate, because the energy of individual layers of identical adjacent atoms is not exactly the same as that in the real compounds.

Based on this assumption, we also obtain the energy of the Bi-Se bond by taking out a single Se atom out of the structure, while excluding the single-layer contribution described in above. The energies of other bonds can be obtained using similar procedure. The results are shown in table. 2.

|  | Layer of Bi | Layer of Se | Layer of Te | |
|---|---|---|---|---|
| Energy (eV) | -40.05 | -15.12 | -23.06 | |
|  | $E_{Bi-Te\_in}$ | $E_{Bi-Te\_out}$ | $E_{Bi-Se\_in}$ | $E_{Bi-Se\_out}$ |
| Energy (eV) | -0.617 | -1.047 | -0.746 | -1.392 |

Tabel.2. the zero-order energies of bonds in $Bi_2Se_xTe_y$.

This results may be compared with other calculations. For example, the first derivative in Fig.3 (a), which is twice the energy difference to replace an outer-layer Se atom in $Bi_2Se_3$ with Te atom, equals to 0.595 eV. The same quantity can be extracted from data in table 2, and the result is 0.597 eV, which is in excellent agreement.

Aside from the linear dependence of total energy with Te concentration, there are also nonlinear behavior in these systems. For example, when exchanging an outer-layer Te atom with an inner-layer Se atom in Bi$_2$Te$_2$Se, the total energy deviate from linear behavior, as is shown in fig.2. Comparing the two curves in fig. 3(a), the nonlinear behavior is stronger for Te doping to the inner layer. We can exclude the linear term in Fig.3 (a) by subtracting the energy of pure Bi$_2$Se$_3$ and Bi$_2$Te$_3$. For example $\Delta E = E - [(1-\frac{x}{9})E_{Bi_2Se_3} + \frac{x}{9})E_{Bi_2Se_2Te}]$ for Bi$_2$Se$_2^{I}$ (Se$_{1-x/9}$Te$_{x/9}$)$^{II}$, and obtain a clear convex function of energy change versus number of doped Te atoms, as shown in fig.3(b).

A natural assumption is that different atom arrangements would introduce different strains into this system, thus affecting the bond length and the bond energy of the system. Comparing the two curves in Fig.3(b) in which Te atoms are doped into either the outer layer or inner layer of Bi$_2$Se$_3$, we find that the energy of doping inner layer Te atoms shows larger nonlinear behavior. Comparing the atomic positions before and after Te doping, we can find a geometrical difference between doping Te atoms to the inner layer and to the outer layers. Fig.4(b) shows the atomic displacements for Te atoms doped in the outer layers, where a large displacement of Te out of the QL, while the displacements of all other atoms are small. In comparison, by doping Te atoms in the inner layer, the displacement of Te is "compressed" by its surrounding atoms, while the surrounding atoms also show relatively large displacements.

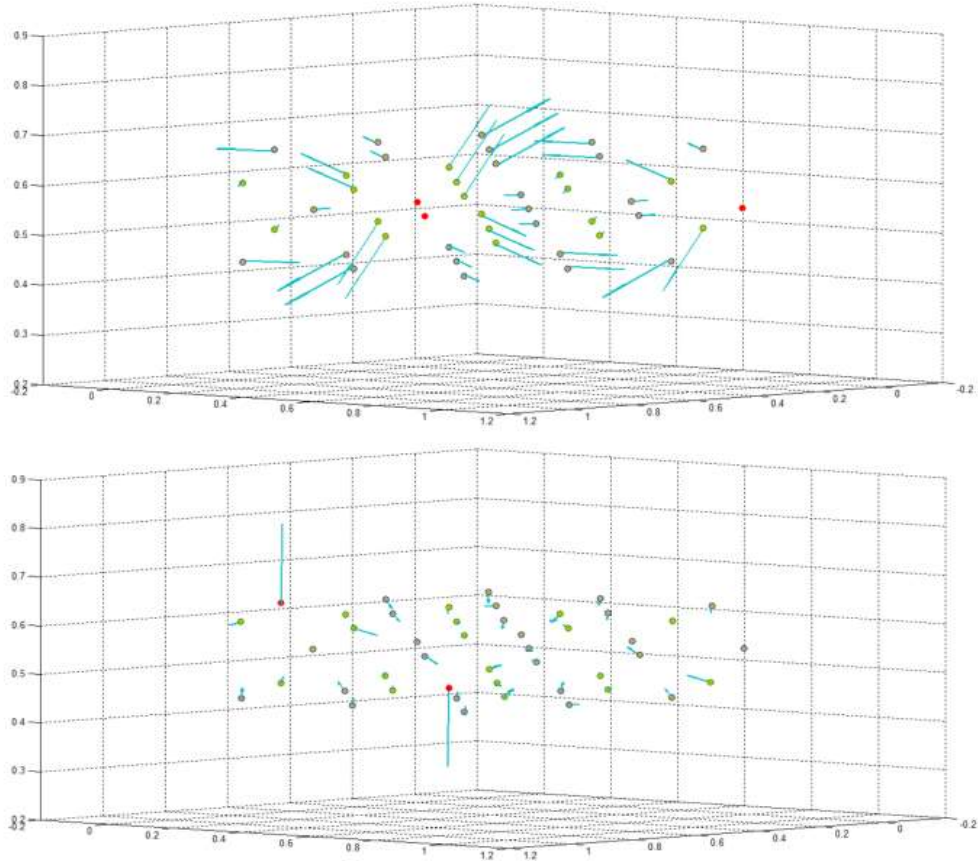

FIG. 4. Displacement of atoms, (a) when Te atoms are doped into the inner layer; (b) when Te atoms are doped into the outer layers of Bi$_2$Se$_3$.

Because the Bi-Te bond length (3.26 Å) is longer than the Bi-Se bond length (3.10Å), the doping of Te behaves like adding a big ball two layers of mattress. If putting the ball on the outer side of one mattress, it will just stay there, making little influence to the rest of the system. However if putting the ball between the two mattresses, it will push up the mattresses and produce a local displacement that contains strain energy. This strain would shorten the Bi-Te bonds, while at the same time enlarge the surrounding Bi-Se bonds. However, if the inner side is completely filled with Te atoms, the strain would disappear again and so is the additional strain energy.

According to the above analysis, we calculate the number of adjacent Te-Se pairs Se. Assuming random distribution of Te and Se atoms, the mean value of adjacent Te-Se pairs is proportional to $x(9-x)$, (x represents the number of doped Te in the inner layer),the result is shown in fig. 3(c). The number of adjacent Te-Se pairs for all the structures in fig. 3(c) matches the additional energies in fig. 3(b) very well, and there is almost perfect correspondence for each data point. Figure 5 shows examples of how to determine the adjacent Te-Se pairs, and a summary of the results is shown in table 3.

As the inner layer Se atoms form a triangular lattice, as shown in fig. 5, the number of adjacent Te-Se pairs equals to the number of triangles that contain both Te and Se. Thus, a simple way to count the total additional energy is to sum the energy of these triangles, if we assume the triangles with pure Te or Se have no strain and will not bring additional energy. Note that the additional energy to dope one Te atom ($x=1$) approximately equals to the energy to dope eight Te atoms ($x=9-1$), we can conclude that the two different types of triangles contain the same additional energy, then it is reasonable to expect that the total additional energy is proportional to the number of adjacent Te-Se pairs, as is shown in fig. 3(e).

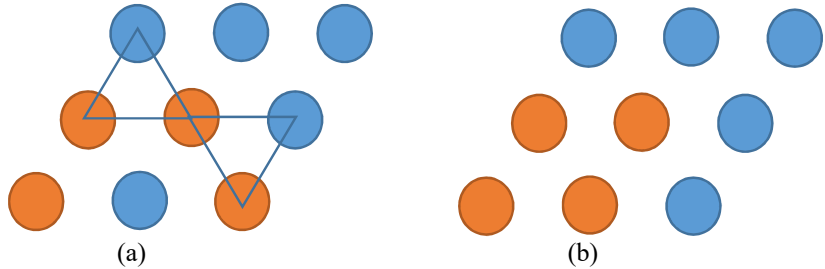

(a)            (b)

FIG. 5. Two structures of the Te/Se layer in $Bi_2Se_3$, with a $3 \times 3$ supercell, where the blue spheres indicate Se atoms and the orange spheres for Te atoms. In structure (a), there are 16 adjacent Te-Se pairs, with 8 triangles containing two Te and one Se atoms, and other 8 triangles containing one Te and two Se atoms. In structure (b), there are 14 adjacent Te-Se pairs, with 4 triangles containing two Te and one Se atoms, and 10 triangles containing one Te and two Se atoms.

| Number of Te | Number of arrangements | Number of adjacent Te-Se pairs |
|---|---|---|
| 1 | 1 | 6 |
| 2 | 2 | 10, 12 |
| 3 | 4 | 12, 14, 18 |
| 4 | 4 | 14, 16, 18 |
| 5 | 4 | 14, 16, 18 |
| 6 | 4 | 12,14,18 |
| 7 | 2 | 10,12 |

| 8 | 1 | 6 |

Table.3 By substituting a given number of Te atoms to a Se layer with $3 \times 3$ supercell, the total number of inequivalent configurations and the number of adjacent Te-Se pairs in the supercell are shown.

In the model above, the two types of triangles have the same energy. This assumption can be further checked. By assuming that the additional energy is proportional to the square of the changes of the Bi-Se and Bi-Te bond lengths, the energy values in fig. 3(b), $E_{change}$, can be expressed as the following:

$$E_{change} = A \sum (L_1 - L_{Bi-})^2 + B \sum (L_2 - L_{Bi-Se})^2], \qquad (4)$$

where $L_2$ stands for the Bi-Te bond length in Te-doped $Bi_2Se_3$, and likewise for $L_1$. If we use the data from table 3 to fit the parameters, the fitting result is shown in fig. 3(d) and 3(f), with $A$ equals to 2.5 eV/ Å$^2$ and $B$ equals to 3.2 eV/ Å$^2$. It can be shown that this result implies the validity of the above assumption.

In addition, based on the conclusion that the additional energy is proportional to the number of adjacent Te-Se pairs, every time we exchange the position of a Te atom and a Se atom, we can lower the total energy of the system by about 36 meV if the number of adjacent Te-Se pairs decreases. This energy is relatively big compared with the room temperature (26 meV). Thus we can perform a Monte-Carlo simulation at low temperature for an initially random arrangement of Te and Se, which can exchange adjacent atoms to lower its energy. Clusters of local Te atoms (assuming the percentage of Te atoms in the layer is less than 50%) will form, as shown in fig. 6, as this type of local Te clusters greatly decreases the number of adjacent Te-Se pairs.

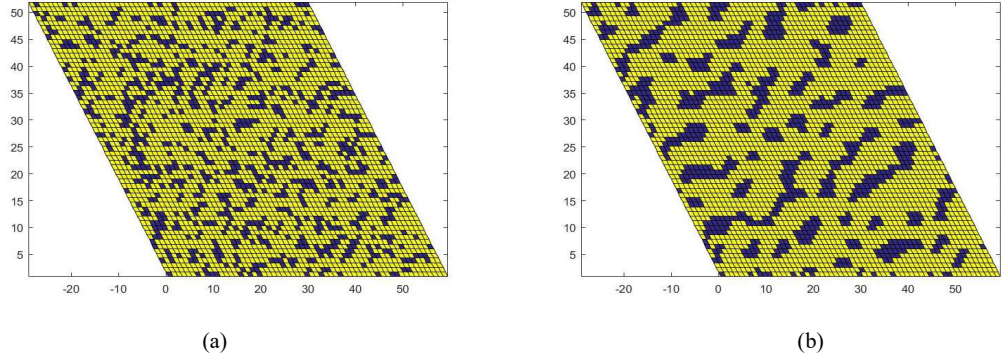

FIG. 6. Monte-Carlo simulation for the arrangement of the Se/Te atoms in a single layer. The simulation is performed at low temperature, and the nearest-neighbor atoms can exchange positions. The yellow dots represent Se atoms while the black ones represent Te atoms, the concentration of Te is set at 25%. (a) The initial random configuration of the Te and Se atoms, (b) the configuration after 160 steps of simulations, in which local clusters of Te atoms have formed.

**Conclusion**

The present study shows that doped Te atoms prefer to be in the outer layer of $Bi_2Se_3$, and a model based on Bi-Te and Bi-Se bond distortion and adjacent Te-Se pairs are developed to explain the phenomenon quantitatively. given a certain doping proportion on a single face of (0,0,1), With this understanding, for any Te doping concentration to $Bi_2Se_3$, we are able to determine the additional energy term and predict the lowest energy arrangements: the fewer adjacent Te-Se pairs, the lower total energy. This result brings a new understanding to substitution of Te atoms into $Bi_2Se_3$, since Te substitution of Se is not completely random but is governed by certain rules. As a result, the properties of this system, such as the electric transport property, might change and show some novel characteristics. Further study of those properties is expected to yield better understanding and applications of these doped topological insulators.

**Acknowledgements**

This work was supported by National Natural Science Foundation of China (Grant Nos. 11474197, 11521404 and U1632272) and National Basic Research Program of China (Grant No. 2013CB921902). Computational resources were supported by Center for High Performance Computing, Shanghai Jiao Tong University.

\* Kaixin Huang and Haotian Zheng contributed equally to this work.
† wdluo@sjtu.edu.cn